# Anti-Stokes Photoluminescence in Monolayer $WSe_2$ Activated by Plasmonic Cavities through Resonant Excitation of Dark Excitons

Niclas S. Mueller[1*], Rakesh Arul[1], Ashley P. Saunders[2], Amalya C. Johnson[2], Ana Sánchez-Iglesias[3], Shu Hu[1], Lukas A. Jakob[1], Jonathan Bar-David[1], Bart de Nijs[1], Luis M. Liz-Marzán[3], Fang Liu[2], Jeremy J. Baumberg[1*]

[1] NanoPhotonics Centre, Cavendish Laboratory, Department of Physics, University of Cambridge, Cambridge CB3 0HE, United Kingdom
[2] Department of Chemistry, Stanford University, Stanford, California 94305, United States
[3] CIC biomaGUNE Basque Research and Technology Alliance (BRTA) Paseo de Miramón 194, Donostia-San Sebastián 20014, Spain
[*]email: nsm44@cam.ac.uk, jjb12@cam.ac.uk

KEYWORDS: plasmonics, 2D material, $WSe_2$, photoluminescence, dark exciton, anti-Stokes PL, NPoM

**Abstract**
**Anti-Stokes photoluminescence (PL) is light emission at a higher photon energy than the excitation, with applications in optical cooling, bioimaging, lasing, and quantum optics. Here, we show how plasmonic nano-cavities activate anti-Stokes PL in $WSe_2$ monolayers through resonant excitation of a dark exciton. The tightly confined plasmonic fields excite the out-of-plane transition dipole of the dark exciton, leading to light emission from the bright exciton at higher energy. Through statistical measurements on hundreds of plasmonic cavities, we show that coupling to the dark exciton is key to achieving a near hundred-fold enhancement of the upconverted PL intensity. This is further corroborated by experiments in which the laser excitation wavelength is tuned across the dark exciton. Finally, we show that an asymmetric nanoparticle shape and precise geometry are key for consistent activation of the dark exciton and efficient PL upconversion. Our work introduces a new excitation channel for anti-Stokes PL in $WSe_2$ and paves the way for large-area substrates providing optical cooling, anti-Stokes lasing, and radiative engineering of excitons.**

**Introduction**

Anti-Stokes photoluminescence (PL) is a process in which light is emitted at a higher energy than the excitation laser by extracting energy from the material. The upconversion occurs through a variety of mechanisms including the absorption of phonons,[1, 2] Auger processes,[3] or multi-photon absorption.[4, 5] This leads to industrially-relevant applications in optical refrigeration,[6] bioimaging,[7] lasing,[8] quantum information,[9] and the detection of infrared light.[10, 11] Excitons in two-dimensional semiconductors are a promising platform for anti-Stokes PL.[2, 5, 12, 13] The reduced dielectric screening and enhanced Coulomb attraction make optical transition dipoles and exciton-phonon coupling an order of magnitude larger than in conventional bulk semiconductors or quantum wells.[14-16] Anti-Stokes PL in transition metal dichalcogenides (TMDs) is thus mostly explained by phonon-assisted processes and the interplay of different excitons.[2, 17, 18]

Efficient anti-Stokes PL requires a condition in which both the excitation laser and the emission are resonant with a material excitation. Initial experiments on anti-Stokes PL in monolayer $WSe_2$ were explained by doubly resonant processes involving charged and neutral excitons,[2, 17] as well as A- and B-excitons.[5] Another possible excitation channel arises from the spin-orbit splitting of the conduction band in TMDs, which leads to the formation of bright and dark excitons with an energy splitting of several tens of meV (Fig. 1a).[18] The dark excitons are either momentum-forbidden (intervalley) or spin-forbidden (intravalley) for excitation at normal incidence.[14, 15] In $WSe_2$ and $WS_2$ they have lower energies than the bright excitons, which leads to strong quenching of the bright exciton emission at low temperature due to fast non-radiative relaxation to the non-emissive dark exciton.[19] The energetic ordering makes dark excitons on the other hand a potential excitation channel for anti-Stokes PL. Excitation with far field radiation is however inefficient, because the dark intravalley exciton has an out-of-plane transition dipole.[20] A promising route to activate the dark exciton is through metallic nanostructures that sustain localized surface plasmons. The tightly confined optical near-fields enable direct coupling to the out-of-plane transition dipole of the dark exciton and enhance the normal Stokes PL emission, which was demonstrated for metallic tips as well as nanoparticle-on-mirror cavities.[21-23]

Here, we show that plasmonic cavities strongly activate anti-Stokes PL in $WSe_2$ through resonant excitation of the dark intravalley exciton. We employ nanoparticle-on-mirror cavities to couple to the out-of-plane transition dipole of the dark exciton as an excitation channel, and enhance the outcoupling through the bright exciton as an emission channel. Through a statistical analysis of hundreds of plasmonic cavities we demonstrate that coupling to the dark exciton is key to enhancing the anti-Stokes PL emission by two orders of magnitude. This is further corroborated by experiments in which we tune the excitation wavelength. Finally, we show that decahedral nanoparticles with asymmetric shapes and precise Au(111) facets are ideal to activate the dark exciton and achieve consistent enhancement of the anti-Stokes PL.

**Results and Discussion**

**Enhancing anti-Stokes PL emission with plasmonic nanoparticle-on-mirror cavities**

We prepare mm-sized monolayers of $WSe_2$ by gold tape exfoliation of bulk van der Waals crystals.[24, 25] The $WSe_2$ monolayers are first exfoliated on a 285 nm $SiO_2$/Si substrate and then transferred onto a template-stripped Au surface with CAB polymer to prevent direct binding with the Au which quenches the PL emission (Fig. S1). Au nanoparticles are then deposited on top of the $WSe_2$ to form nanoparticle-on-mirror (NPoM) plasmonic cavities (Fig. 1b).[26, 27] The NPoM cavities confine light to the nm gap between the nanoparticle bottom facet and the Au mirror enclosing the $WSe_2$ monolayer, which leads to a near $10^4$- fold enhancement of the local light intensity (Fig. 1b, right). The plasmonic near fields have both in-plane and out-of-plane components which allow coupling to the bright exciton ($X_B$) with an in-plane transition dipole as well as the dark exciton ($X_D$) with an out-of-plane transition dipole (Fig. 1a, b).[23]

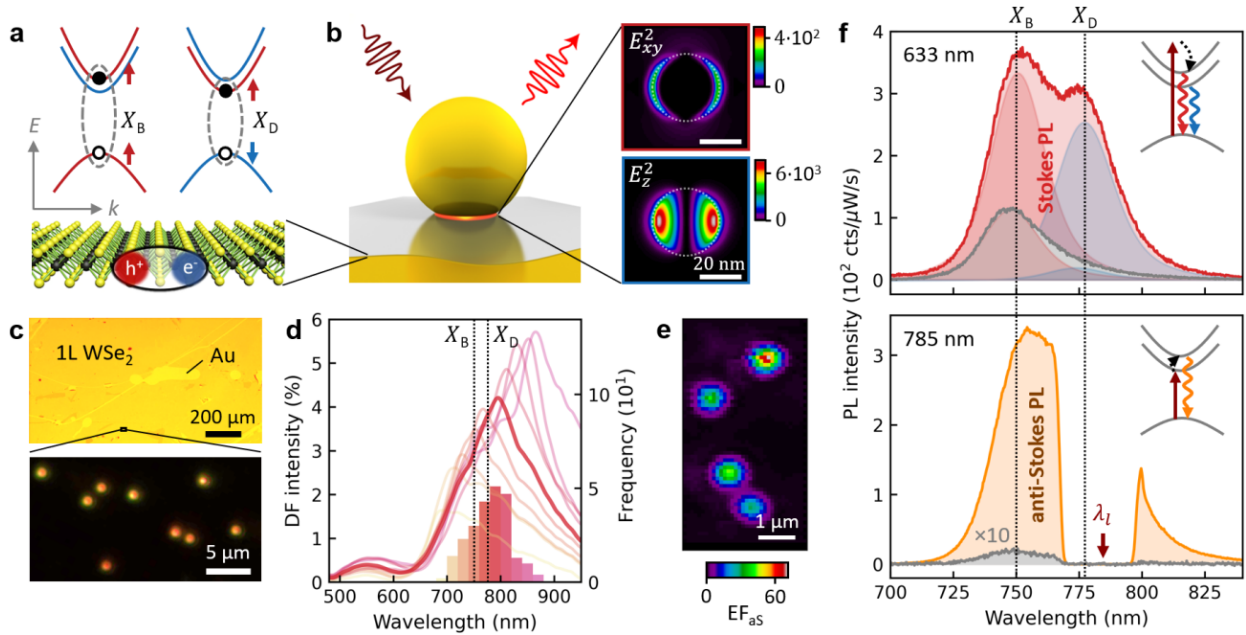

**Figure 1: Activation of dark exciton and anti-Stokes PL in WSe₂ by plasmonic cavities.** (a) Band-structure of spin-allowed bright ($X_B$) and spin-forbidden dark ($X_D$) excitons in WSe₂. (b) Nanoparticle-on-mirror (NPoM) plasmonic cavity with WSe₂ defining a gap between the nanoparticle bottom facet and Au mirror. Simulated field intensity enhancement in WSe₂ is shown on right, with in-plane $E_{xy}^2$ and out-of-plane $E_z^2$ field enhancement. Dotted lines show nanoparticle bottom facet. (c) Optical bright-field micrograph of large-area monolayer WSe₂ on Au (top), and dark-field micrograph of NPoM cavities on WSe₂ (bottom). (d) Histogram of dark field scattering spectra of plasmonic NPoM cavities from measurements on 252 cavities. (e) Map of anti-Stokes PL by scanning $\lambda_l = 785$ nm excitation laser across four cavities with precise decahedral nanoparticle shapes (see Fig. 4 below). (f) Stokes PL spectrum with laser excitation wavelength $\lambda_l = 633$ nm (top, red) and anti-Stokes PL spectrum with $\lambda_l = 785$ nm (bottom, orange). Reference spectra recorded beside plasmonic cavities are shown in grey. Peak components from fits of bright (red) and dark (blue) excitons are shown for Stokes PL. Insets show energy diagrams of the Stokes (red, blue) and anti-Stokes PL (orange) emission, with excitation laser (dark red) and internal energy conversion (black dotted). Dotted lines show energies of bright and dark excitons.

Our simple approach to sample preparation allows the simultaneous fabrication of thousands of NPoM cavities encapsulating WSe₂ monolayers, easily located from their dark-field scattering pattern in an optical microscope (Fig. 1c). Using particle recognition algorithms and a fully automated setup we measure the optical spectra of hundreds of plasmonic cavities. The dark-field scattering spectra show a broad plasmonic mode around 790 nm that overlaps with the expected emission wavelengths of the dark and bright excitons of WSe₂ (Fig. 1d). Variations of the confined plasmon wavelength partly originate from different nanoparticle shapes and geometries of the bottom facet (see below).[28]

The PL emission of WSe₂ is strongly enhanced when centering the laser measurement spot on an NPoM cavity (Fig. 1e,f). We choose two excitation laser wavelengths $\lambda_l$ to yield Stokes PL ($\lambda_l = 633$ nm) and anti-Stokes PL ($\lambda_l$ = 785 nm). The Stokes PL is enhanced by a factor of 4.3 compared to the signal recorded beside the plasmonic cavity (Fig. 1f, top). Most strikingly, a new emission peak appears at 777 nm that is shifted by ~50 meV from the bright exciton $X_B$ at 750 nm. We attribute this additional peak to PL emission from the dark intravalley exciton $X_D$, as also reported in a recent work on NPoM cavities (Fig. 1f, top, inset).[23] The dark exciton has an out-of-plane transition dipole that couples selectively to the optical field components in the plasmonic cavity that point normal to the WSe$_2$ monolayer.[21-23] As the plasmonic near-fields in the nm gap of NPoM cavities are primarily polarized out-of-plane (Fig. 1b, right), the NPoM geometry is ideal to activate the dark exciton. Coupling to momentum-indirect dark excitons is less efficient because of their in-plane transition dipole, and momentum matching with plasmonic nanostructures would require a field confinement on the sub-nm scale.[29] Weak emission from the dark intravalley exciton is already visible in the reference spectrum recorded beside the NPoM cavity because we collect the PL emission with a high numerical aperture objective (Fig. 1f, grey).[20] Comparison to the reference spectrum allows us to estimate an enhancement factor $EF_D$ = 12 for the dark exciton, while the bright exciton is only enhanced by $EF_B$ = 3 (see peak components in Fig. 1f). As exciton-phonon coupling also contributes to the asymmetric lineshape of the reference spectrum, the enhancement of the dark exciton is probably even much larger.[30]

The enhancement of the PL emission becomes an order of magnitude larger when moving to an excitation wavelength of $\lambda_l$ = 785 nm that leads to anti-Stokes PL emission, as this is near-resonant with the dark exciton (Fig. 1f, bottom). The enhancement $EF_{aS}$ = 170 is so large that the anti-Stokes PL for $\lambda_l$ = 785 nm becomes comparable to the Stokes emission for $\lambda_l$ = 633 nm, while the reference spectrum recorded beside the NPoM is near undetectable. The PL enhancement is tightly localized to the position of the plasmonic cavities (Fig. 1e). From line scans of the excitation laser spot across single NPoM cavities we extract a spatial FWHM of $EF_{aS}$ ~ 460±30 nm, close to the diffraction limited laser spot size $\lambda_l/2NA$ = 440 nm (Fig. S2). This is very different from previous work, where the PL signal was only slightly reduced even μm distances away from the plasmonic cavities.[13] When considering the size mismatch of the laser excitation spot with the area of the nanoparticle bottom facet where enhancement occurs, we estimate local enhancement factors of $EF_{aS,loc}(785\ nm) \approx 9 \cdot 10^4$, $EF_{D,loc}(633\ nm) \approx 4 \cdot 10^3$, and $EF_{B,loc}(633\ nm) \approx 9 \cdot 10^2$ for the spectra in Fig. 1f (SI section S4). The difference in enhancement at 785 nm and 633 nm excitation cannot be explained by the wavelength dependence of the plasmonic near field intensity when only considering the bright exciton (SI section S6). This hints to the dark exciton as an additional excitation channel for anti-Stokes PL (Fig. 1f, bottom, inset).

To test this hypothesis, we characterize the PL enhancement of 252 NPoM cavities (Fig. 2). For each plasmonic cavity we correlate the enhancement of the anti-Stokes PL with both dark and bright excitons in the Stokes PL emission (Fig. 2a). Almost all NPoM cavities enhance the PL emission, with the dark exciton enhancement $\langle EF_D \rangle = 7$ on average larger than that of the bright exciton $\langle EF_B \rangle = 3$. The enhancement factors, however, vary by an order of magnitude and not all NPoM cavities activate the dark exciton. The variation in anti-Stokes PL enhancement is even larger, ranging from $EF_{aS}$ = 1 to >200, with an average $\langle EF_{aS} \rangle$ = 20. The average enhancement is in excellent agreement with finite-difference time-domain (FDTD) simulations which predict $EF_{aS}$ = 21 (SI section S6). To better understand the large spread in enhancements, we compare the average PL spectra of NPoMs that activate the dark exciton ($EF_D/EF_B \geq 2$, Fig. 2b left) with NPoMs that do not activate the dark exciton ($EF_D/EF_B < 2$, Fig. 2b right). Indeed, when the dark exciton is activated the average enhancement of the anti-Stokes PL is larger $\langle EF_{aS} \rangle = 65$ than for NPoM cavities that do not couple to the dark exciton $\langle EF_{aS} \rangle = 21$. This hints to a mechanism in which (i) the 785 nm laser excites the dark exciton, (ii) energy is extracted from the material e.g. through phonon-assisted processes, and (iii) anti-Stokes emission finally occurs via the bright exciton (Fig. 2c). Plasmon-mediated coupling to the dark exciton is thus key to enhance the anti-Stokes PL. The activation of the dark exciton $EF_D/EF_B$, however, varies by one order of magnitude and only a subset (30%) of the NPoM cavities efficiently activate the dark exciton, which explains the large spread of $EF_{aS}$ (Fig. 2d, S6). We discuss how to improve this below.

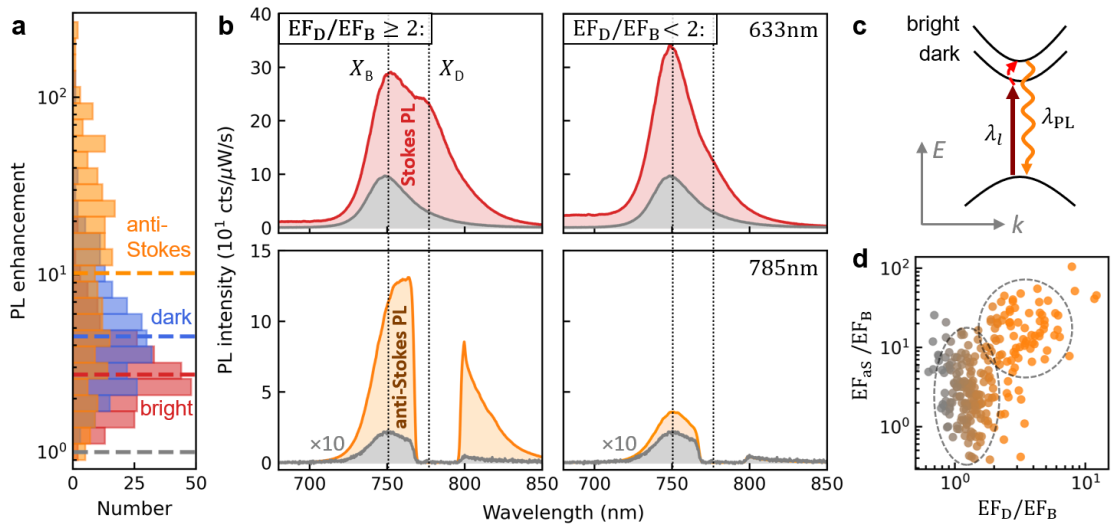

**Figure 2: Correlation between dark exciton activation and anti-Stokes PL enhancement.** (a) Histogram of PL enhancement factors from measurements on 252 NPoM cavities. Enhancement of anti-Stokes PL ($EF_{aS}$, orange) from $\lambda_l = 785$ nm pumping, and enhancement of dark ($EF_D$, blue) and bright ($EF_B$, red) excitons from $\lambda_l = 633$ nm pumping. Dashed lines show median enhancement factors. (b) Average Stokes (top) and anti-Stokes (bottom) PL spectra for NPoMs that activate the dark exciton ($\langle EF_{aS} \rangle = 65$, left) vs NPoMs that do not activate the dark exciton ($\langle EF_{aS} \rangle = 21$, right). (c) Anti-Stokes PL process via resonant excitation of dark exciton; red dashed arrow shows coupling to phonons. (d) Relative strength of anti-Stokes PL enhancement $EF_{aS}$ *vs* activation of dark exciton $EF_D$, each normalised to bright exciton. Dashed lines show clustering of data points for NPoMs that couple (orange) or not (grey) to the dark exciton.

To better understand the excitation mechanism of anti-Stokes PL, we sweep the laser excitation wavelength from $\lambda_l$ = 730-805 nm across the resonances of the dark and bright excitons and measure excitation-emission profiles (Fig. 3). The emission evolves from Stokes PL at $\lambda_l$ = 730 nm to anti-Stokes PL at $\lambda_l$ = 805 nm (Fig. 3a). Without plasmonic cavities, the Stokes emission is largest when exciting blue-detuned from the bright exciton, while anti-Stokes emission is largest when resonantly exciting the bright exciton (Fig. 3b, top). This is expected, as the bright exciton is the only excitation channel. The excitation-emission profile changes profoundly in the presence of a plasmonic NPoM cavity (Fig. 3b, bottom). An additional resonance appears at excitation wavelengths red-detuned from the bright exciton, and the anti-Stokes emission is largest for $\lambda_l$ close to the dark exciton. This is also consistently seen in the excitation-emission profiles of other NPoM cavities (Fig. 3c) and shows that the dark exciton serves as an excitation channel for anti-Stokes PL.

The anti-Stokes PL is most efficient for excitation wavelengths between the dark and bright excitons, as the upconversion process requires energy to be extracted from the material, which becomes less efficient with increasing energy difference of excitation and emission. The upconversion can occur either through the coupling to thermally populated phonons, or electron-electron scattering.[2, 18, 19] Nonlinear excitation, e.g. through two-photon absorption,[31] can be excluded here, as we observe a near-linear power dependence of the anti-Stokes PL at these ~10 µW CW excitation powers (SI Section S3). The energy shift between $\lambda_l$ for largest anti-Stokes PL and the bright exciton (~35 meV) indeed matches the $A_1'$ phonon of $WSe_2$ (250 cm$^{-1}$ = 31 meV), which hints to a phonon-mediated upconversion, as suggested previously.[2] The overall excitation wavelength dependence is then determined by an interplay of plasmon-enhanced excitation of the dark exciton and upconversion through phonons and other mechanisms. To extract the role of the plasmonic cavities we determine the anti-Stokes PL enhancement for each excitation wavelength (Fig. 3d). In contrast to the overall PL emission, the enhancement factor peaks at excitation wavelengths slightly red detuned from the dark exciton at $\lambda_l \approx 785$ nm, as used here in all other measurements. The aS-PL is enhanced by ≳100 for wavelengths beyond the dark exciton that match the plasmon resonances of the NPoM cavities.

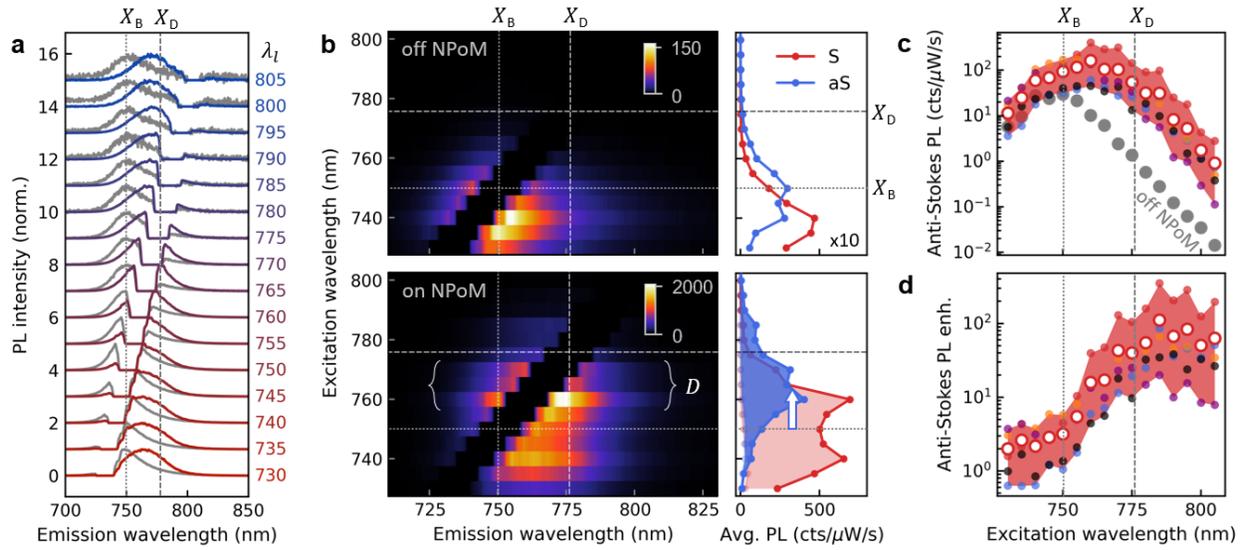

**Figure 3: Excitation wavelength dependence.** (a) Normalized PL spectra enhanced by NPoM cavity (coloured) and beside cavity (grey) while tuning excitation wavelength $\lambda_l$ from 730 nm (bottom) to 805 nm (top). (b) Excitation emission map of PL spectra measured beside cavity (top) and enhanced by NPoM cavity (bottom). Colour bars are in counts/µW/s. Panels on right show average signal of anti-Stokes (aS, blue) and Stokes PL (S, red). Label '$D$' refers to new resonance activated by dark exciton. (c) Anti-Stokes PL signal and (d) enhancement of five NPoM cavities (small colored dots), their average (red circles), and reference measurement beside cavities (grey). Dashed lines show wavelengths of dark and bright excitons.

Finally, we address the large variation in anti-Stokes PL enhancement factors (see Figs. 2a,d). A similar spread in enhancement was also observed in surface-enhanced Raman scattering from NPoM cavities and was attributed to variations in nanoparticle shape and size (Fig. 4a, left).[32] The plasmonic resonances of NPoM cavities mainly depend on the geometry of the nanoparticle's bottom facet,[28] which varies considerably for the commercial nanoparticles used here. We therefore employ decahedral nanoparticles with precise geometry and triangular Au(111) facets (Fig. 4a, right).[33] These were recently demonstrated to yield more consistent SERS enhancement and plasmonic modes when assembled as nanodecahedra-on-mirror (NDoM) cavities (Fig. 4b).[34] Indeed, we observe a three-fold narrower distribution of plasmon resonance wavelengths than for NPoM cavities (compare Figs. 4c and 1d). The dominant NDoM plasmonic mode centred around 780 nm for this size is chosen to spectrally overlap with the dark exciton of $WSe_2$.

The plasmonic resonances are spectrally narrower than for NPoM cavities and arise from higher-order plasmonic modes that are located under the bottom facet of the nanoparticles (Fig. S9).[34]

The precise geometry of the NDoM cavities also leads to a much more consistent enhancement of the PL emission of $WSe_2$ (Fig. 4e). Compared to NPoMs the distribution of enhancement factors is two-fold narrower for Stokes PL and ten-fold for anti-Stokes PL. The NDoM cavities consistently activate the dark exciton with $\langle EF_D \rangle = 7.5$ and $\langle EF_B \rangle = 2$, while 90% of the cavities provide $EF_D/EF_B > 2$. This leads to an enhancement of the anti-Stokes PL $\langle EF_{aS} \rangle = 44$ that is on average larger than for NPoM cavities which give $\langle EF_{aS} \rangle = 20$, while the enhancement is similar to that of NPoMs activating the dark exciton (Fig. S6b). The measured enhancement is in good agreement with FDTD simulations for the Stokes PL but underestimated for anti-Stokes PL (SI section S6-3). Possible reasons are cancellations of different emission channels in our simulations, a different dark to bright conversion on and beside the nanoparticle, and effects of exciton diffusion that are not included in our simulations. We attribute the remaining spread in enhancement factors to intrinsic spatial variations of the $WSe_2$ PL emission that occur across the substrate (Fig. S5), and the distribution of plasmon resonance wavelengths. Indeed, when sorting the enhancement factors by the distribution of plasmon resonance wavelengths, we find that $EF_D$ and $EF_{aS}$ are largest when the plasmon resonance wavelength matches the dark exciton, which is the case here for almost all NDoMs (Fig. 4d, blue and orange). The bright exciton is on the other hand detuned from the plasmon resonances leading to a smaller $EF_B$ than for NPoMs (Fig. 4d, e red). Since the most efficient anti-Stokes PL requires bright and dark exciton to be both resonant with plasmonic modes, this gives room for further optimization.

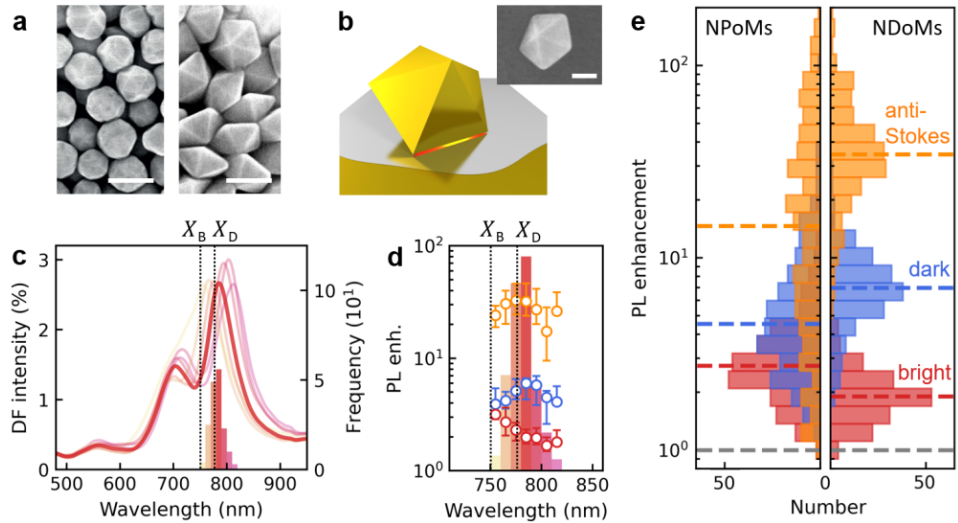

**Figure 4: Consistent enhancement with nano-decahedral cavities.** (a) SEM image of nanoparticles used for NPoM cavities above (left, adapted from [32]) *vs* decahedral nanoparticles with precise shape and size (right). Scale bars are 100 nm. (b) Sketch of nanodecahedra-on-mirror (NDoM) plasmonic cavity. Inset shows SEM image of NDoM (scale bar 50 nm). (c) Histogram of dark field (DF) scattering spectra of 173 NDoM cavities. (d) Median of $EF_{aS}$ (orange), $EF_D$ (blue) and $EF_B$ (red) *vs* DF peaks in (c). Error bars are from first to third quartile. (e) Histograms of enhancement factors for NPoMs (left) vs NDoMs (right). Dashed lines show median EFs.

In conclusion, we showed that plasmonic coupling to the dark exciton in WSe$_2$ activates a highly efficient excitation channel for anti-Stokes PL. This leads to a near hundred-fold enhancement of PL upconversion that occurs within tightly spatially confined fields of plasmonic cavities. Compared to previous work, our large-area samples allowed us to analyse the optical spectra of hundreds of plasmonic cavities and correlate the enhancement factors of different emission channels. We showed that an asymmetric nanoparticle shape and precise geometry is key for coupling to the dark exciton and activating anti-Stokes PL. Our approach may be extended to localized excitons at excitation wavelengths below the dark exciton, that are activated through local strain or defects, and to other 2D luminescent materials.[35, 36] The simplicity and scalability of our sample fabrication paves the way for large-area substrates for PL upconversion, which may find applications in optical refrigeration of 2D materials, anti-Stokes lasing, and radiative engineering of excitons in the future.

**Methods**

**Sample preparation.**

*Exfoliation of WSe$_2$:* WSe$_2$ monolayers are exfoliated with the gold tape exfoliation method from bulk single crystals (HQgraphene).[24] In brief, a 100 nm Au layer is evaporated on a Si wafer with an ebeam evaporator. The Au surface is spin coated with a layer of polyvinylpyrrolidone (PVP) as a protective layer against the further contaminations from processing. The PVP/Au stack is stripped from the Si surface with a thermal release tape (Nitto), and is released onto a 285 nm SiO$_2$/Si substrate upon heating. The PVP protection layer is further dissolved with water, and the Au is etched with KI/I$_2$ solution. The obtained monolayers are transferred onto a templated stripped Au substrate with CAB polymer, and cleaned with ethyl estate.[37]

*Preparation of NPoM cavities:* For the preparation of NPoM cavities, 80 nm Au nanoparticles (BBI Solutions, mean diameter 77 - 85 nm) were mixed with 0.1 M NaNO$_3$ (10:1), drop casted on a selected area of the sample, and washed off with DI water after 10 s. Nanoparticles that remain on the substrate form NPoM cavities and were identified with a dark field microscope and spectrometer. Following similar steps as for the preparation of NPoMs, the decahedral nanoparticles were drop casted onto the WSe$_2$ substrate to form plasmonic NDoM cavities.[34]

*Synthesis of gold decahedra:* Decahedral Au nanoparticles were synthesized using seed-mediated growth following previous protocols.[33, 34] In brief, penta-twinned gold seeds were used to grow decahedral nanoparticles in a solution containing benzyldimethylhexadecyl-ammonium chloride (BDAC, 50 mL, 100 mM), gold (III) chloride trihydrate (HAuCl$_4$, 0.5 mL, 50 mM, ≥ 99.9%) and ascorbic acid (AA, 0.375 mL, 100 mM, ≥ 99%) to an edge length of 41 nm. Gold decahedra with an edge length 70 nm were then synthesized by adding 41 nm gold decahedra seeds (0.5 mL, 5 mM) to a growth solution containing BDAC (50 mL, 15 mM), HAuCl$_4$ (0.25 mL, 50 mM) and AA (0.19 mL, 100 mM). The gold decahedra solution was centrifuged at the different growth steps to remove excess reactants and dispersed in aqueous CTAB solution (10 mM). Finally, the decahedra were functionalized with citrate through subsequent centrifugation and redispersion in poly (sodium 4-styrenesulfonate) (Na-PSS, 50 mL, 0.15 % wt, M$_w$~70000 g/mol), and finally sodium citrate (50 mL, 5 mM, ≥ 99%).

**Photoluminescence and dark field spectroscopy.**

Optical spectra were recorded with custom-built setups that consist of a dark field microscope with a motorized stage and spectrometers. The setups are fully automated to record the optical spectra of hundreds of plasmonic cavities. We used particle recognition algorithms to locate NPoM cavities by their dark field scattering pattern and centre them to the measurement spot. Dark field (DF) scattering spectra were recorded by illuminating the sample with incoherent white light (halogen lamp) through an Olympus MPLFLN 100x DF objective with 0.9 NA and recording the scattered light with a fibre-coupled QE Pro spectrometer (Ocean Optics). For each NPoM cavity an automatic scan through the focal depth is performed, and the depth-dependent DF intensity is used to correct the spectrum from chromatic aberrations.

For PL spectroscopy we used either CW single-frequency diode lasers with $\lambda_l = 632.8$ nm (Integrated Optics, MatchBox) and $\lambda_l = 785$ nm (Thorlabs, LP785-SF100), or a tuneable cw Ti:Sapphire Laser (SolsTiS, M Squared Lasers). The lasers were focussed to a diffraction-limited laser spot through the 100x DF objective and the laser power on the sample was kept below 20 µW, unless stated otherwise. The emitted PL was collected with the same objective and guided to a grating spectrometer (Andor Kymera or Shamrock, 150 l/mm grating) with a CCD camera (Andor Newton EMCCD). The spectrometer slit width and readout lines on the CCD were limited to detect PL from a single nanoparticle. For automated measurements comparing 633 nm and 785 nm excitation, the two lasers were combined with a dichroic mirror and subsequently switched on. Pairs of notch filters for the two laser wavelengths were subsequently moved into the detection path with automated sliders. For measurements with tuneable laser excitation, we used angle-tuneable laser clean up filters and notch filters. All spectra were dark count subtracted with spectra recorded on the bare Au substrate. The datasets for statistical measurements on NPoM and NDoM cavities were pre-screened, and data were discarded from further analysis if the excitation laser was defocussed or shifted from the nanoparticle, or if the nanoparticle was beside $WSe_2$. Reference spectra were recorded 2 µm beside each plasmonic cavity and the average spectra from all positions were used to calculate enhancement factors. All measurements were conducted at room temperature.

**Scanning electron microscopy.**

For SEM characterization, Au nanoparticles were cleaned three times with ultrapure water by centrifugation (3000 rpm) to remove the extra surfactant as much as possible. 3 µl concentrated Au nanoparticle solution was dropped on a silicon substrate which was then dried in a vacuum chamber. The SEM measurement was performed on Hitachi S4800 with a 10 keV accelerating voltage.

**Finite-difference time-domain simulations.**

To simulate the electric field enhancement of NPoM cavities, we used the software package Lumerical FDTD Solutions from Ansys. We identified a geometry of an NPoM cavity that is representative for our experiments by reproducing the experimental scattering spectra (Fig. S7). The nanostructure was illuminated from the top with a Gaussian beam source with 0.9 NA and the electric field intensity in the NPoM gap was recorded with an electric field monitor. The electric fields were normalized by the total electric fields without the plasmonic nanostructure to calculate enhancement factors. All further details of the simulations are provided in the Supporting Information (SI section S6).

## ASSOCIATED CONTENT

**Supporting Information**

PL spectra of WSe$_2$ on Au vs glass substrate; Spatial localization of PL enhancement; Excitation power dependence of anti-Stokes PL; Estimate of local experimental enhancement factors by NPoM cavities; Statistics of dark exciton activation by plasmonic cavities and peak fits; FDTD simulation of PL enhancement.

## AUTHOR INFORMATION

**Corresponding Authors**

* Dr Niclas S Mueller, nsm44@cam.ac.uk
* Prof Jeremy J Baumberg, jjb12@cam.ac.uk* Dr Niclas S Mueller, nsm44@cam.ac.uk
* Prof Jeremy J Baumberg, jjb12@cam.ac.uk

**Author contributions**

N.S.M. and J.J.B. conceived the project. N.S.M., R.A., L.J., J.B.-D., and B.d.N. designed the optical setups, N.S.M. and R.A. conducted the optical experiments, and S.H. electron microscopy. A.P.S., A.C.J. and F.L. prepared the WSe$_2$ samples. A.S.-I. and L.M.L.-M. synthesized the decahedral nanoparticles. The manuscript was written with contributions from all authors.

**Competing interests**

The authors declare no competing financial interest.

## ACKNOWLEDGEMENTS

The authors acknowledge funding from the EPSRC (EP/L027151/1 and EP/R013012/1), and the ERC (883703 PICOFORCE, 861950 POSEIDON). B.d.N. acknowledges support from the Winton Programme for the Physics of Sustainability and from Royal Society University Research Fellowship URF\R1\211162. L.M.L-M. acknowledges funding from the Spanish Ministerio de Ciencia e Innovacion, MCIN/AEI/10.13039/501100011033 (Grant PID2020-117779RB-100). N.S.M. acknowledges support from the German National Academy of Sciences Leopoldina. R.A. acknowledges support from the Rutherford Foundation of the Royal Society Te Apārangi of New Zealand, the Winton Programme for the Physics of Sustainability, and Trinity College Cambridge. L.A.J. acknowledges support from the Cambridge Commonwealth, European & International Trust and EPSRC award 2275079. J.B.D acknowledges support from the Blavatnik fellowship. F.L. acknowledges support from Terman Fellowship and startup funds from Department of Chemistry at Stanford University. We thank Angela Demetriadou for helpful discussions.